\documentclass[floatfix,twocolumn,showpacs,preprintnumbers,amsmath,amssymb]{revtex4}

\usepackage[latin1]{inputenc}
\usepackage[dvips]{graphicx}
\graphicspath{{./pics/}}

\renewcommand {\vec}[1]{{\mathbf{#1}}}

\newlength{\figurewidth}

\begin{document}
\pacs{68.65.Hb,81.16.Dn,81.16.Rf}
\title{Strain-controlled correlation effects in self-assembled quantum 
dot stacks}

\author{R.~Kunert}
\author{E.~Sch{\"o}ll}
\email{schoell@physik.tu-berlin.de} 
\affiliation{Institut für Theoretische Physik, Technische Universität 
Berlin, D-10623 Berlin, Germany}

\begin{abstract}
We show that elastic interactions of an array of self-assembled quantum 
dots in a parent material matrix are markedly distinct from the elastic 
field created by a single point defect,  
and can explain the observed abrupt correlation--anticorrelation 
transition in semiconductor quantum dot stacks. Finite volume 
effects of the quantum dots are shown to lead to sharper transitions.
Our analysis also predicts the
inclination angle under which the alignment in successive quantum dot 
layers occurs in dependence on the material anisotropy. 
\end{abstract}

\maketitle


The self-assembled growth of semiconductor quantum dots (QDs) 
is widely used as an efficient tool for fabricating 
nanostructures with important potential applications in
opto- and nanoelectronics  \cite{STA04,SHC04}.
The process of self-assembly is based on the 
formation of coherent localized structures confined in three
spatial directions, i.e. QDs, on a flat, uniform two-dimensional (2D) wetting 
layer in highly strained materials, using the 
Stranski-Krastanov growth mode.
Multilayer systems in the form of vertical stacks are used to improve the 
lateral ordering of QD arrays, and minimize the 
size and shape fluctuations of the QDs. 
By tuning the spacer thickness a certain desired lateral
spacing of QDs can be obtained, and one can switch between 
several types of stacking modes \cite{XIE95,HOL99,SPR00,WAN04}. 
Theoretically, several approaches have been made to understand the
new features of QD stacks. Elasticity theory has proven to be very
useful, not only for understanding the growth of QD systems consisting
of one layer \cite{SHC99}, in particular when combined with kinetic Monte Carlo 
simulations \cite{MEI01,MEI01b,MEI03a}, but also for giving insight 
into the growth of
stacked QD systems \cite{XIE95,TER96a,SHC98,HOL99,MEI03,HEI03}. 
Our understanding of the vertical and lateral correlation 
and anticorrelation 
properties of such stacked arrays of QDs is, however, still incomplete.
Recently, clear experimental evidence of an abrupt transition between
vertically aligned (correlated) and antialigned (anticorrelated) QDs in
the InGaAs/GaAs material system in dependence upon the spacer layer
thickness has been presented \cite{WAN04}, but this transition could not 
be explained by the anisotropic strain field of a single QD 
\cite{HOL99,SPR00};
and the angle against the vertical under which the alignment in successive
QD layers is observed seriously disagrees with the theoretically predicted tilt
angle for the minima of the strain \cite{SPR00}.

In this report we present a three-dimensional model of the 
strain field of a stacked array of quantum dots, 
taking fully into account the anisotropy of the 
material system and the 3D shape of the QDs. This is
an extension of the model proposed by 
Shchukin \textit{et al.} \cite{SHC98}. 
While other theoretical descriptions \cite{SHC98,HOL99}
using a similar Green's tensor  approach \cite{POR77}
have employed strictly periodic arrays and point-like or stripe-like 
approximations of the QDs,
we have implemented arbitrary QD shapes and configurations.
Within our approach we are able to explain 
both the experimentally investigated 
tilt angles and positions in an anticorrelated stacked array of QDs 
in the InGaAs/GaAs material system \cite{WAN04} and describe the
transition from correlated to anticorrelated growth.


We model the multi-sheet array of QDs by inclusions of material A
in a matrix of material B.
Starting from elasticity theory for anisotropic materials
\cite{KHA83,SHC98}, we solve 
the equilibrium equation
$\nabla_j \left( \lambda_{ijkl} \nabla_k u_l (\vec{r}) \right) = 
  \nabla_j (\sigma_{ij}^{(0)} \vartheta(\vec{r}))$
with the elastic moduli $\lambda_{ijkl}$, 
the elastic displacement field  $u_l$, 
the characteristic shape function $\vartheta(\vec{r})$ of the QDs 
($\vartheta(\vec{r})= 1$ if $\vec{r}$ is inside an
inclusion and zero elsewhere),
and the stress tensor $\sigma^{(0)}_{ij}$ which is connected to
the stress-free strain tensor $\varepsilon^{(0)}_{ij}$
by $\sigma^{(0)}_{ij} = \lambda_{ijkl}\varepsilon^{(0)}_{kl}$ \cite{KHA83}. 
Here Einstein's summation convention is used.
Since the volume of the inclusions is assumed to be small compared to
the parent material, the homogeneous moduli approximation is used. 
The problem is solved under stress-free boundary conditions on the
surface.

In general, the elastic energy of interacting inclusions is given by \cite{KHA83}
\begin{equation}
  \label{eq:el_generall}
  E_{el} = \frac{1}{2} \int d^3\vec{r} 
  \sigma_{ij}^{(0)} \vartheta(\vec{r}) 
  \left( \varepsilon_{ij}^{(0)} - \varepsilon_{ij}(\vec{r})\right).
\end{equation} 
Introducing the in-plane coordinates $\vec{r}_\parallel$ and the vertical 
coordinate $z$, 
one can express the elastic interaction energy 
per unit surface area in the growth plane at
$z = z_0$ in terms of the static Green's tensor as \cite{SHC98}
\begin{equation}
  \label{eq:e_elast_simple}
  E(\vec{r}_\parallel; z=z_0 ) =
  h^{(S)}
  \int d^2\vec{r'}_\parallel \int_{-\infty}^{z_0} d z'
  \vartheta( \vec{r'}_\parallel; z') 
  W(\vec{r}_\parallel -  \vec{r}'_\parallel; z_0-z')   
\end{equation}
where the integral is over the buried QDs with $z'\leq z_0$, 
$h^{(S)}$ is the height of one single inclusion on the surface,
and the interaction term
\begin{equation}
  \label{eq:integral_first}
  \begin{split}
   W(\vec{r}_\parallel - \vec{r}'_\parallel; z_0 - z') = 
   \int \frac{d^2 \vec{k}_\parallel}{(2\pi)^2} 
   e^{ i\vec{k}_\parallel (\vec{r}_\parallel-\vec{r}'_\parallel) } \\
   \times \left\{ \sigma^{(0)}_{ij} 
     \left[ \nabla_j \nabla'_{m} 
       \tilde{G_{il}} (\vec{k}_\parallel;z_0,z_0 - z')
     \right] 
     \sigma^{(0)}_{lm} 
   \right\} 
   \end{split}
\end{equation}
is related to the Fourier transform of the static Green's tensor $\tilde{G_{il}}$,
which can be computed numerically for elastically anisotropic cubic crystals, 
for details see \cite{POR77,KHA83,IPA98,SHC98}.


\begin{figure}[htb]
  \includegraphics[width=\figurewidth]{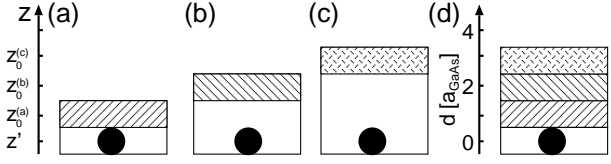}
  \caption{Schematic representation of the dependence of the surface elastic energy 
    upon the spacer thickness. The cross-sections of samples with 
    different spacer thickness $d$ 
    (a)-(c) are combined into a single cross-section (d).
    }
  \label{fig:sketch}
\end{figure}
\begin{figure}[htb]
  \includegraphics[width=\figurewidth]{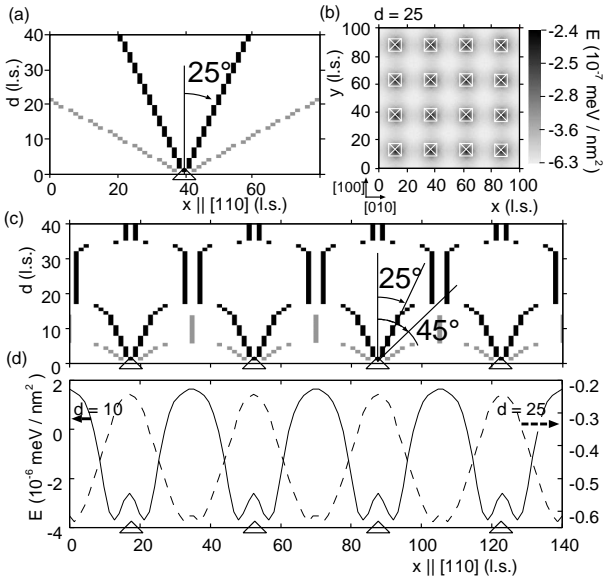}
  \caption{Elastic strain field of 
    one single point-like inclusion (a) 
    and a periodic array of point defects (b)-(d) in a GaAs matrix.
    (a) and (c) show the minima (black) and maxima (grey) of the elastic energy
    $E$ per unit area in the [110] direction of the relaxed surface vs. 
    spacer thickness $d$.
    The density plot (b) depicts $E$ in the 
    [001]-[010] surface plane at $d=25$.      
    (d) $E$ for $d=10$ and $d=25$. 
    The positions
    of the QDs are indicated by triangles (a),(c),(d) and
    by boxes (b), respectively.
    All lengths are in units of the lattice constant
    of the parent material $a_{GaAs} = 0.565$~nm (monolayer width). 
    }
  \label{fig:point-array}
\end{figure}
\begin{figure}[htb]
  \includegraphics[width=\figurewidth]{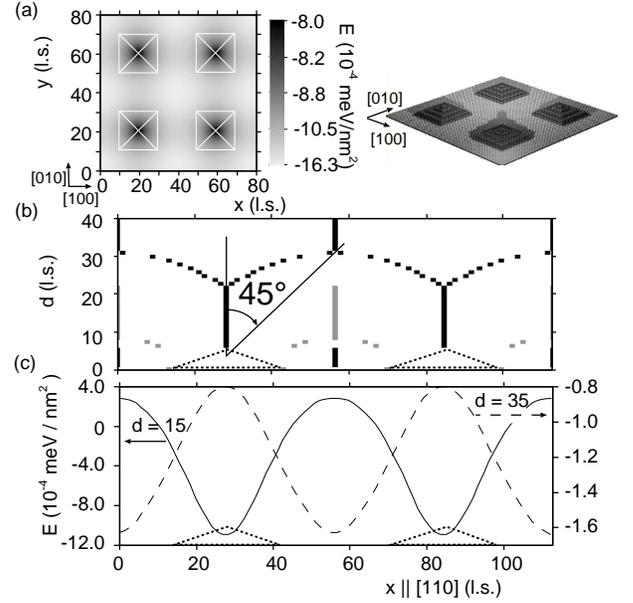}
  \caption{Elastic strain field of 
    a periodic array of pyramids in a GaAs matrix.
    (a) density plot of the elastic energy per unit area $E$ in the 
    [001]-[010] surface plane at $d=35$ (inset: 3D view) (b) minima (black) 
    and maxima (grey) of $E$
    in the [110] direction of the relaxed surface vs. spacer 
    thickness $d$. (c) $E$ for $d=15$ and $d=35$.
    The positions and sizes of the QDs are indicated by boxes (a) and 
    dotted triangles (b),(c).
    }
  \label{fig:pyr-array}
\end{figure}
We have computed the elastic energy 
in the relaxed surface at $z = z_0$ for a given array of buried QDs
at $z = z'$ in dependence on the spacer thickness $d = z_0 - z'$. 
The results are combined into a single cross-sectional plot as schematically
illustrated in Fig.~\ref{fig:sketch}.  
Periodic boundary conditions in the lateral direction are used. 
Fig.~\ref{fig:point-array}(a) shows the minima (black) and the maxima (grey)
of the elastic interaction energy $E$ in the relaxed surface for a single buried point-like
QD. The minima and maxima occur in the [110] direction at angle $\alpha=25^\circ$ 
and $\tilde{\alpha}=60^\circ$ against the vertical, 
and correspond to maximum tensile strain 
($E<0$, black) and maximum compressive strain ($E>0$, grey), respectively. 
Regions of tensile strain give rise to an attractive potential and act
as preferred nucleation
centers for QDs, whereas regions of compressive strain form
repulsive potentials \cite{PEN01}.

As can be seen, 
our model is in good agreement with previous theoretically predicted values
of the angle $\alpha$ 
in which the minima of the elastic energy in the
[110]-[001]-plane occur \cite{HOL99}.  
Computing the inclination angle $\alpha$ for other material
systems, we find that $\alpha$ increases 
with increasing anisotropy ratio $A = 2c_{44} /(c_{11}-c_{12})$ 
of the parent material, where $c_{11}, c_{12}, c_{44}$ are the elastic moduli in 
Voigt notation. 
We obtain $\alpha = 19^\circ$ for Si ($A = 1.56$), $\alpha = 25^\circ$ for GaAs 
($A = 1.83$), and $\alpha = 33^\circ$ for ZnSe ($A=2.04$), in agreement with
\cite{HOL99}. However, these results contradict the experimental observation of much
larger angles \cite{WAN04}.

The situation changes drastically if we consider an array of QDs
(Fig.~\ref{fig:point-array}(b)-(d)). 
We use a quadratic array of
$4\times4$ QDs with a lateral spacing of $l = 25$~lattice sites (l.s.). 
In a neighborhood of approximately $l/2$ around each QD the elastic
properties are governed only by this single point defect. Outside this
region the strain fields are overlapping, which leads to new effects
(Fig.~\ref{fig:point-array}(c)). 
Starting with a spacer thickness of $d\approx 14$~l.s., the inclination angle 
$\alpha$ increases more and more until at 
$d = 17$~l.s. the two minima induced by two neighboring QDs 
meet and form a flat double minimum exactly in the middle between these QDs.
With further increasing spacer thickness the position of the double minimum 
remains stable and unchanged for a considerable range of $d$-values. Finally,
at $d = 31$~l.s the minimum starts moving back to a position vertically above the
QDs. 

The transition of the minima from a position close to vertically above the QDs 
to in between the QDs is also visible in the elastic energy profiles at $d=10$ l.s.
and $d=25$ l.s. in Fig.~\ref{fig:point-array}(d). This indicates a transition 
from correlated to anticorrelated growth.
From Fig.~\ref{fig:point-array}(c) we infer an inclination angle for 
alignment of QDs of about $\alpha = 45^\circ$ for 
GaAs which is in reasonable agreement with the experimentally observed
value of $50^\circ$ \cite{WAN04}.
The small remaining discrepancy might result from several subtle effects.
First, in our model the material of the QDs is completely neglected.
Second, 
stacking of the QDs might also be influenced by modulations in chemical 
composition or be due to morphological changes in the spacer layer. 

For different materials, again the inclination angle $\alpha$ 
increases with increasing anisotropy, as in the case of 
one single QD:  $\alpha = 43^\circ$ for Si, $\alpha = 45^\circ$ for GaAs, and 
$\alpha = 47^\circ$ for ZnSe.

We have also studied the influence of the shape and volume of the QDs.
In Fig.~\ref{fig:pyr-array} we consider a $2\times 2$ array of pyramids, 
each with a baselength of $20\times 20$~l.s. and a height of 5~l.s
positioned with a lateral distance of $l= 40$~l.s. 
Compared with an array of defects, there are significant differences. 
The inclination angle $\alpha$ 
remains unchanged. But the finite volume of the QDs results in a large
range of spacer thickness $d$ where the minimum energy 
is directly located vertically above the buried structures 
(Fig.~\ref{fig:pyr-array}~(b)). At a spacer  
thickness of about four times the height of the 
pyramids the energy minimum splits and moves towards 
the anticorrelated positions between the QDs. Due to this finite size 
effect, the transition from the correlated to the anticorrelated regime 
happens much more abruptly than in an array of point defects  
(Fig.~\ref{fig:point-array}(c)). Such an abrupt transition is indeed 
observed
experimentally \cite{WAN04}. Also the double minimum of the energy profile is 
replaced by a single minimum (Fig.~\ref{fig:pyr-array}(c)) .
Note that the transition occurs at larger spacer thickness $d\approx32$ 
for the array of QDs with a larger lateral distance of $l=40$ l.s., and this
ratio should scale for arrays with even larger lateral distance at fixed
angle $\alpha$. This is consistent with the experiment \cite{WAN04} where the
transition was observed at a spacer thickness of 150 monolayers for an
in-plane nearest-neighbor distance of 80--100 nm.

In conclusion, we have shown that elastic
interactions of an array of QDs in a 
parent material matrix explain the observed correlation--anticorrelation 
transition, in contrast 
to calculations which take into account only a single 
point defect. The observed material trend of increasing inclination angles 
of the vertical stacking mode for larger anisotropy is also corroborated.
The abrupt transition has been attributed to the finite volume of the QDs, and the 
reclining angle under which the anticorrelated QDs align is in good quantitative
agreement with the experiment on self-organized InGaAs/GaAs quantum dot stacks.

The algorithm introduced in this paper 
allows for fast computation of the elastic 
interaction energy of nano\-structures with 
arbitrary shape and spatial arrangement in three dimensions, and the only 
parameters are the elastic moduli of the considered 
anisotropic cubic parent material. Thus it is appropriate for 
implementation into kinetic Monte Carlo simulations \cite{MEI03,MEI03a} 
which might be used for a detailed investigation of the growth kinetics
of self-assembled QD stacks in future research.

The authors would like to thank V.A.~Shchukin for helpful discussions.
This work was supported by Deutsche Forschungsgemeinschaft
in the framework of Sfb 296.


\end{document}